# 2P-MED: BUILDING A PERSONALIZATION PLATFORM FOR MEDIATION SYSTEMS

IMANE ZAOUI[*], DALILA CHIADMI, LAILA BENHLIMA

Department of Computer Science,
Mohammadia School of Engineers(EMI), Mohammad Vth University -Agdal
BP. 765 Av. Ibn Sina Agdal Rabat Morocco. http://www.emi.ac.ma
imane.zaoui@gmail.com, chiadmi@emi.ac.ma, benhlima@emi.ac.ma

Abstract:
Nowadays, with the increasing number of integrated data sources, there is a real trend to personalize mediation systems to improve user satisfaction. To make these systems user sensitive, we propose a personalization platform called 2P-Med. 2P-Med allows personalizing any mediation system used in any domain following a cyclic process. The process includes building and managing adequate user profiles and sources profiles, content and quality matching, source selection, adapting the mediator responses to user preferences and handling user feedbacks. In this paper, we describe 2P-Med architecture and highlight its main functionalities. We also illustrate the operation of the platform through personalizing source selection in a travel planning assistant.

*Keywords: Personalization; user profile; source selection; mediation systems.*

1. Introduction

More and more domains like digital libraries, e-commerce, e-government, e-health, information retrieving and web search engines make use of data integrating technologies to offer value added services and richer information. To integrate virtually disparate and heterogeneous data sources, mediation systems have appeared as a challenging but nevertheless efficient solution [Wiederhold, (1992)]. Indeed, Mediators resolve interoperability problems [Shen, (2006)], transcend geographical spread and deal with the autonomy of the distributed data sources. The purpose of a mediator is to provide a unified and logical view on the top of existing data sources. Users access the view via a unified query interface that offers location, model, and interface transparency. Unfortunately, with the increasing number of data sources especially in the context of the web, getting a satisfying response becomes a hard task. Thus, one of the fundamental aspects of user interaction in mediation systems is user satisfaction. This is clearly mentioned by industry reports who stress that although data integration initiatives succeed in achieving a common technology platform, they are rejected by the user communities due to the information overload or the presence of poor data quality [Peralta, (2006)]. Indeed, the mediator gives the same response for the same query although user interests are different. For example for the query "I look for travel offers to Morocco", the mediator returns a listing of travel offers without any distinction of the purpose of the trip or kind of accommodation since these information are not clearly mentioned in the request. This scenario reflects the inability of the mediator to distinguish between users and to all satisfy their needs and preferences. A solution for these limitations is personalization. Personalization is the capability to provide content and services that are tailored to individual users based on knowledge about their preferences and behaviors [Hagen, (1999)]. By using personalization techniques, the mediator is able to deliver more suitable answers that meet user needs and respect his quality requirements which significantly improve user satisfaction.
In this paper, we present 2P-Med, a **P**ersonalization **P**latform for **Med**iation systems. 2P-Med is a user sensitive platform that could be plugged on the top of any mediation system. Its main functionalities are recognizing users through user profiles, modeling integrated data sources via sources profiles, and adapting dynamically the mediator answers to each user profile.
The reminder of this paper is structured as follows. Section 2 will present a brief background of the personalization process. In section 3, a prototype implementation of 2P-Med will be described and in section 4, a practical experimentation example will demonstrate the benefit of our method. We conclude in section 5.

2. Personalization background

Personalization is a discipline that deals with adapting the content, the structure and/ or presentation of system responses to each individual user's characteristics, usage behavior and/or usage environment [Kobsa *et al.*., (2001)]. Personalized services are usually delivered to users through user-adaptive systems that took place between providers and users. These personalization systems follows an iterative process based on understand, deliver, and measure cycle [Adomavicius and Tuzhilin., (2005)]. We extend this cycle to the following personalization process.

### 2.1. *The personalization process*

Based on Adomavicius works [Adomavicius and Tuzhilin., (2005)], we propose a personalization process composed of five stages which are user identification, source identification, query execution, impact analyzing and dynamic adaptation.

- **Stage 1 and stage 2 (User identification and source identification)**: The identification requires accurate modeling of both users and sources knowledge. A model, also called profile, contains explicit assumptions on all aspects that may be relevant for the dialogue behavior of the system [Kobsa and Wahlster, (1989)]. Several approaches for modeling exist depending on the application nature and the system goals [Gowan, (2003)][Amato and Starracia, (1999)][Kostadinov, (2007)]. We propose a generic and multi-dimensional profile that could be used in a large variety of domains and applications. In the multi-dimensional approach, the profile is composed of several dimensions. Each dimension contains couples of attributes and values. The attributes are key words or parameters, whereas values are weights and thresholds.
- **Stage 3 (Query execution):** At this level, three fundamental functions are executed based on the user profile. User queries are analyzed and enriched; relevant sources are selected and the mediator responses are customized to respect user preferences and needs, for example, documents are reordered according to their relevancy.
- **Stage 4 (Impact analyzing):** This stage is about determining the impact of using the personalization tool on user satisfaction. This analysis is based on mining users' feedbacks and measuring system performances. The Impact analysis is then exploited to insure a dynamic adaptation of the system.
- **Stage 5 (Dynamic adaptation):** Dynamic adaptation is based on learning methods and concerns especially updating the profiles, and refining the mediator responses. This is performed via executing the personalization process as a cyclic procedure. Thus, the adaptation insures a better understanding of user's characteristics for getting more satisfying responses.

Due to the lack of space, we present in this paper only user and source identification and the personalized source selection procedure.

### 2.2. *User identification*

User identification is based on building the user profile which includes modeling and data acquisition. Several models have been proposed [Zemirli, (2008)][Kostadinov, (2007)][Tanudjaja and Mui, (2002)][Amato and Starracia, (1999)], but they depend on the application domain and do not consider all aspects that characterize the user. The model we propose improves the existing models. First, it gives a complete and temporal view of the user in both short and long terms. Second, it is possible to distinguish between the static and the dynamic user characteristics. Finally, it allows also following user behavior evolution across multiple interaction sessions. Our model organizes the user profile in two entities: the *persistent profile* and the *session profile* presented by the class diagram in Fig. 1.

- **Persistent profile:** contains general characteristics of a user which don't change for a long period. It includes five dimensions: the *personal identity* (name, age, address, etc); the *domains of interest* which group all attributes and preferences related to the general information needs of a given user, domains of interests are formulated using the vector model of Salton [Salton *et al*., (1975)]; the *general expected quality*; the *security data* and the *interaction history*. This persistent profile could be used in building user communities which is a key issue for mediation systems providing recommendation and collaborative filtering services [Nguyen, (2006)].
- **Session profile:** This is a short term profile describing the user during one session. The session profile contains four dimensions: the *user context* which informs about the location, time, used devices, user roles etc; the *user goals* which relates to the specific user domains of interest during a particular session; the *user quality preferences* that covers required quality parameters and values in the current session. The session profile helps to catch the evolution of user characteristics and needs through multiple sessions. This is necessary in updating the knowledge about the user and adapting the system responses.

To build the persistent profile, we adopt an explicit approach [Gowan, (2003)] where the user fills his personal identity, domains of interest, security data and expected quality dimensions through an interactive interface when he create his account. The session profile on the other side is acquired semi-automatically. Indeed, it is possible for the user to fill his context, goals and quality preferences through the querying interface.

Furthermore, the system is able to extract this information from user queries. Finally, the interaction history dimension is captured automatically across multiple sessions.

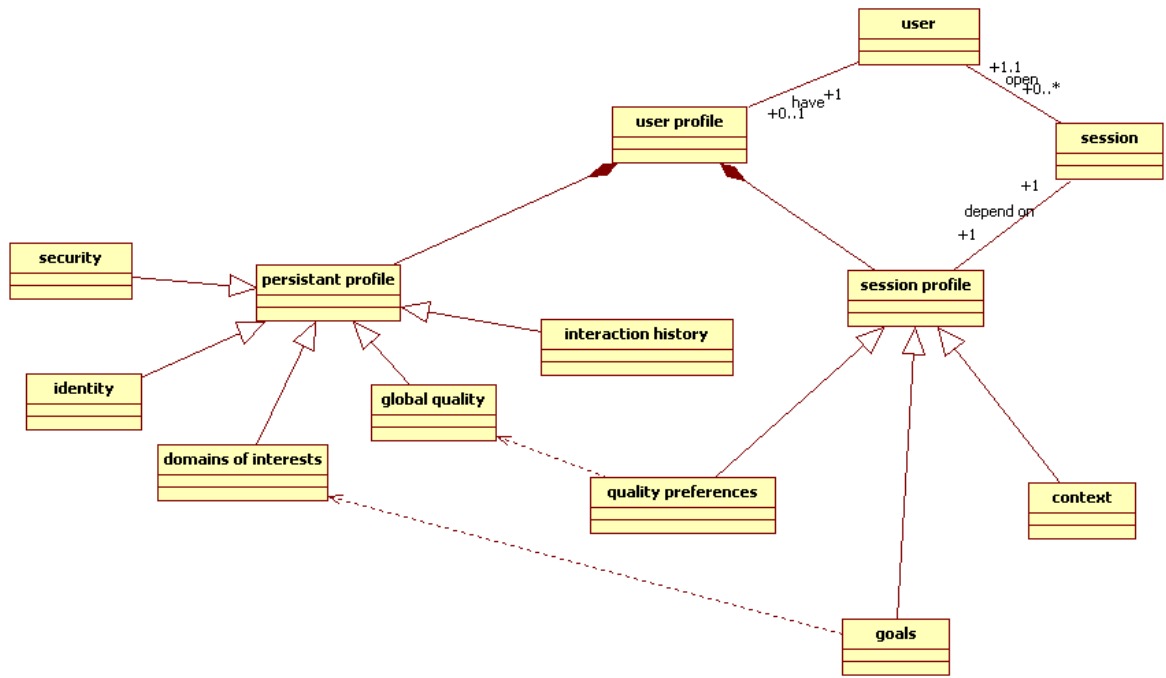

Fig. 1: Class diagram of the user profile

### 2.3. *Source identification*

Sources in mediation systems are heterogeneous, distributed and autonomous. To describe them, local schemas and ontologies are generally used. But these descriptions are not sufficient to give details about the content and the quality criteria of the integrated data sources. For this reason, we refine the sources descriptions by building a source profile that is independent from the application domains. The source profile contains a variety of information including source location, identity, owner, content, quality criteria and so on. A brief description of our source profile dimensions is given bellow. For details refer to [Zaoui *et al*, (2010)].

- **Source identity** gives the source identity using the following attributes: Id, name, URL, port, owner, size, principal languages and principle types of content.
- **Source content** represents the most important topics treated by the sources. This information is available in the form of weighted vector of key words. The key words are the main concepts treated by the source. They are extracted using a term extractor tool [Nazarenko and zargayouna, (2009)]. The weights are calculated using the well known TF*IDF formula[Salton *et al*., (1975)].
- **Source ontology** represents the sources concepts with their semantic relationship.
- **Source quality** describes the main quality characteristics of the source in terms of quality criteria such as freshness, popularity, response time, etc. for more details about measuring source quality criteria; refer to [Wadjinny *et al*., (2011)].

To build our source profiles, we exploit information given by the sources as a meta-data description. For unavailable information, for example for some quality parameters, we use sample queries which helps discovering all source profile dimensions.

### 2.4. *Source selection*

During the query execution stage in the personalization process, a source selection is performed to select the most relevant sources according to the user profile. The source selection procedure given in Fig.2 is based on matching user session profile and sources profiles at two levels. First, a content matching selects the sources that meet user goals. Then, a quality matching refines the selection and ranks them according to user quality preferences.

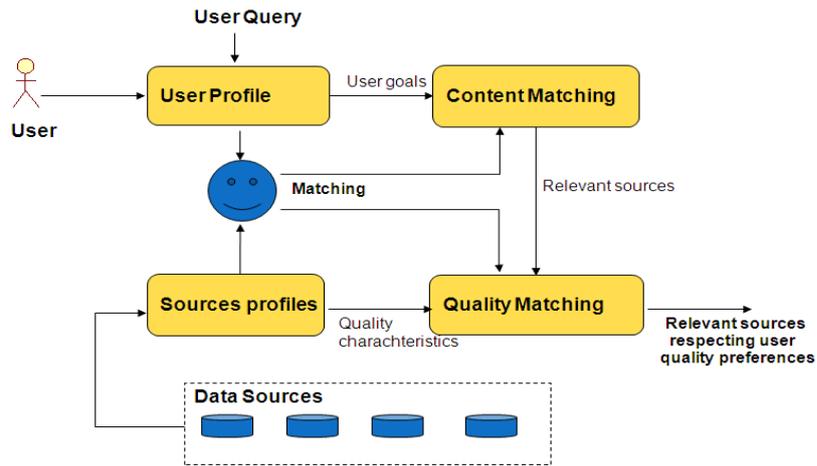

Fig. 2: Personalized source selection procedure based on content matching and quality matching

### 2.5. *Content matching*

The content matching is based on calculating a similarity score between the user goals vector and the sources content vectors. There are a variety of similarity measures in the literature. We choose to adapt the most popular ones which are, the Cosine angle (see Eq. (2)), Jaccard coefficient (see Eq. (3)), and Dice Coefficient (see Eq.(4)). For comparisons between these similarity measures, reader could refer to [Egghe, (2010)].

To straight the importance of terms in the vector, it is preferable to use the cosine angle measure. If we want to measure similarity as a proportion of common terms and non common ones, we use the Jaccard coefficient. To emphasize the importance of common terms instead of non common ones, it is better to use the Dice coefficient. Since similarity between the user goals and the sources content vectors depends on the number of common terms, on their weights, since the common terms are more appreciated, we propose in Eq. (1) to calculate the similarity as a function of the Cosine, Jaccard and Dice coefficients.

### 2.6. *Measuring similarity*

Given $UG(U_i)=(tu_{i1},wu_{i1}; tu_{i2},wu_{i2};....tu_{in},wu_{in})$ the user goals vector and $SC(S_i)=(cs_{i1},ws_{i1}; cs_{i2},ws_{i2};....cs_{in},ws_{in})$, the source content dimension of source $S_j$. We note $Sim(U_i,S_j)$ the similarity score between $UG(U_i)$ and $SC(S_j)$. It is given by the formula:

$$Sim(U_i,S_j)= a\left(\frac{\alpha}{\sqrt{bc}} + \frac{\beta}{b+c-a} + \frac{2*\gamma}{b+c}\right) \qquad (1)$$

where $a=\sum_{k=1}^{n} wu_{ik}\cdot ws_{jk}$ ; $b=\sum_{k=1}^{n} wu_{ik}^2$ ; $c=\sum_{k=1}^{n} ws_{jk}^2$ and $\alpha,\beta,\gamma$ are fitness parameters $\in [0,1]$

*Remark:* some special cases

- If $(\alpha,\beta,\gamma) = (1,0,0)$ then $Sim(U_i,S_j) = Cosine = \frac{\sum_{k=1}^{n} wu_{ik}\cdot ws_{jk}}{\sqrt{\sum_{k=1}^{n} wu_{ik}^2 * \sum_{k=1}^{n} ws_{jk}^2}}$ (2)

- If $(\alpha,\beta,\gamma) = (0,1,0)$ then $Sim(U_i,S_j) = Jaccard = \frac{\sum_{k=1}^{n} wu_{ik}\cdot ws_{jk}}{\sum_{k=1}^{n} wu_{ik}^2 + \sum_{k=1}^{n} wu_{ik}^2 - \sum_{k=1}^{n} wu_{ik}\cdot ws_{jk}}$ (3)

- If $(\alpha,\beta,\gamma) = (0,0,1)$ then $Sim(U_i,S_j) = Dice = \frac{2*\sum_{k=1}^{n} wu_{ik}\cdot ws_{jk}}{\sum_{k=1}^{n} wu_{ik}^2 + \sum_{k=1}^{n} ws_{jk}^2}$ (4)

- If $(\alpha,\beta,\gamma) = \left(\frac{1}{3},\frac{1}{3},\frac{1}{3}\right)$ then $Sim(U_i,S_j) = AVG(Cosine, Jaccard, Dice)$ (5)

Comparing similarity scores allows ranking the sources from the most relevant to the less relevant. User defines also a threshold which helps reducing the set of relevant sources. Using Definition 1, only sources having a similarity score higher than this threshold will be selected.

**Definition 1:** *Given a user threshold Rt and a similarity score Sim($U_i$,$S_j$), the source $S_j$ is relevant if Sim($U_i$,$S_j$)*100 ≥ Rt.*

### 2.7. *Quality matching*

After content matching, a quality matching is performed to return the most relevant sources in terms of quality characteristics. Since the quality of the sources involved is measured through a multitude of criteria, the quality matching could be studied as a multi-attribute decision making problem (MDMP). In [Wadjinny *et al*., (2011)] we propose an algorithm based on the SAW method (Simple Additive Weighting) [Hwang and Yoon, (1981)] which is the most simple but nevertheless a good decision making procedure. The algorithm returns in four stages, an ordered set of only relevant sources that respect user quality preferences.

- **Stage 1**: Select all sources respecting the user quality preferences for mandatory and desirable criteria,
- **Stage 2**: Build the Scaled Decision Matrix given by:

$$M'=[v'_{ij}]_{(n*m)} \text{ where } v'_{ij}= \frac{v_{ij}-min_i(v_{ij})}{max_i(v_i)-min_i(v_i)} \qquad (6)$$

where $v_{ij}$ is the value of the j$^{th}$ quality criterion measured on source $S_i$

- **Stage 3**: Calculate the SAW Score Given by:

$$Score(S_i)=\sum_{j=1}^{m} v'_{ij} * wq_j \qquad (7)$$

where $wq_j$ is the weight of the j$^{th}$ quality criterion measured on source $S_i$

- **Stage 4**: Rank sources according to the SAW Scores

### 3. Building 2P-Med prototype

In traditional mediation systems, users suffer from information overload. For the same query, they got the same responses even if their needs, their context and their preferences are different. The second lack is a long response time due to the mediation process in particular with a multitude of data sources. It has been proved in [Wadjinny, (2010)] that the increasing number of data sources leads to a bottleneck and degrades considerably the mediator performances. 2P-Med platform overcome these limitations. It helps reducing the amount of integrated data through the personalized source selection procedure and insures adapting the mediator responses to users' profiles. In this section, we start from describing the architecture of 2P-Med prototype.

### 3.1. *The architecture of 2P-Med prototype*

To execute the personalization process presented previously, we build 2P-Med prototype on four software components which are: 1) *user manager,* manages the users through their profiles; 2) *source manager,* manages the sources through their profiles; 3) *personalization core* insure personalizing the mediator responses and *4) knowledge base*, stores the users profiles and the sources profiles. Each component contains several modules (see Fig.3). The modular design of the platform allows its extensibility and facilitates its management. Indeed, it is easy to add new modules that insure new functionalities. 2P-Med prototype components are described in the following.

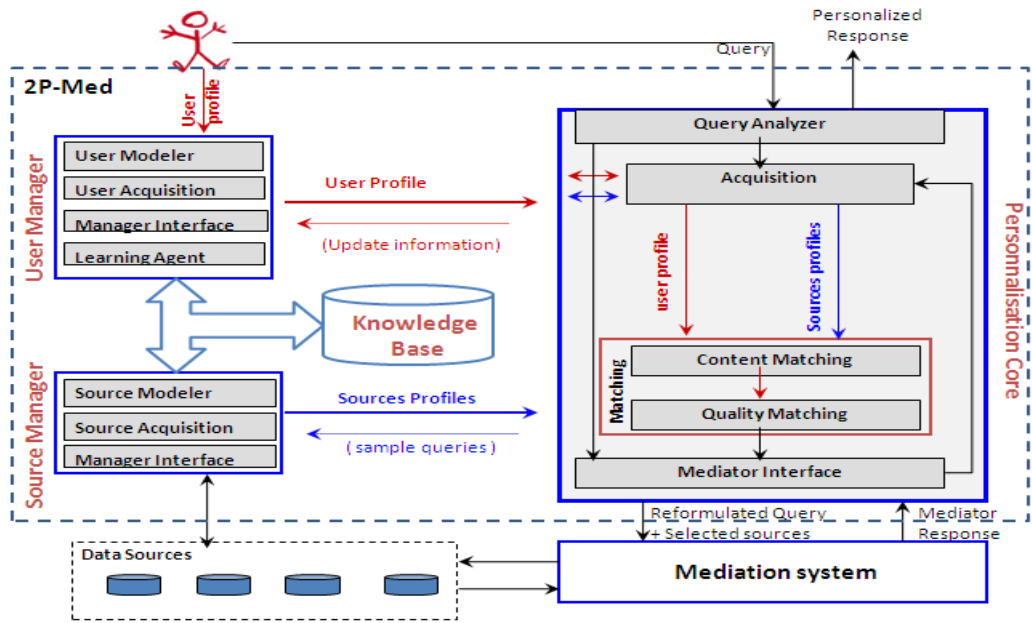

Fig. 3: 2P-Med architecture

### 3.2. *User manager component*

This component involves both the users and the mediator manager according to the use case diagram of Fig.4. Its main functionality is managing users through their profiles. This component contains four modules. The *user modeler* module, instantiates the appropriate user profile according to user needs and the domain of application. For example, a user profile for an e-government application could be composed of only personal identity and context dimensions where as a user profile for digital libraries may contain the domains of interests, the goals and the quality preferences dimensions. The *user acquisition* module fills the user profile information using adequate forms and captures the updated data. The *manager interface* allows users to modify their profiles or delete them. This module provides also a back office interface to the mediator manager to create and manage user accounts. The last module is the *learning agent* which analyzes the user feedbacks, his behavior and also his interaction history to update the user profile. This module also builds community profiles by clustering similar profiles and matches a user profile to the appropriate community profile. Community profiles are used to offer recommendation and collaborative services.

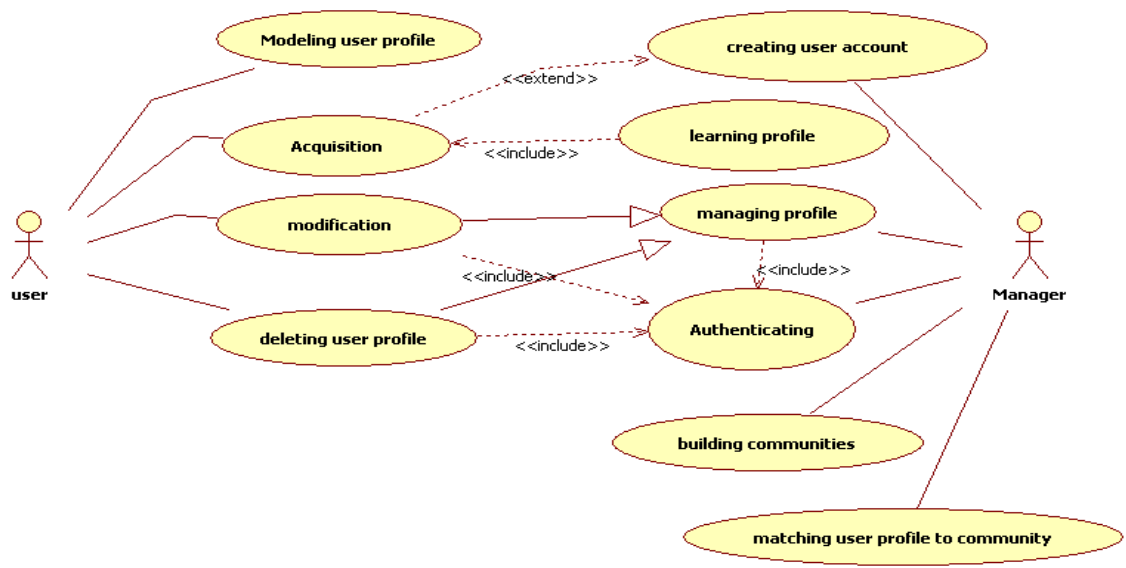

Fig. 4: Use case diagram of the user manager component

### 3.3. *Source manager component*

This component manages the profiles of available data sources. It is composed of three modules. The *source modeler* module builds the source profile depending on the domain of application and the targeted services. The *source acquisition* module fills the source module with the appropriate information which is directly available as metadata or automatically learned using sample queries. The sample queries are sent to the sources to capture their properties such as some quality parameters, content and security rules. The *manager interface* gives interfaces to create a new source profile, to modify and delete profile information. The source manager component use case is given in Fig. 5.

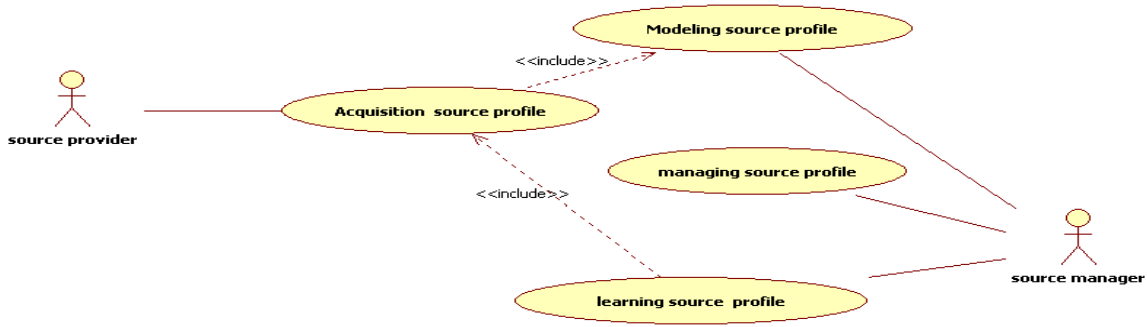

Fig. 5: Use case diagram of the source manager component

### 3.4. *Personalization core component*

This is the core of 2P-Med platform. It involves the user profile, the user query and the sources profiles to personalize the mediator answer. This component is built on five modules presented as actors in the use case of Fig. 6. The main functionalities are analyzing user queries via the *query analyzer module*. This analysis allows extracting the user goals and preferences for each interaction session if the user didn't fill his session profile. Then, the *acquisition module* extracts the adequate information from the user profile and the sources profiles. It also handles the update information and sends it to the user manager component. User profile and sources profiles data are then sent to the *matching module*. The matching is performed in two times. First, a content matching gives the relevant sources according to user goals, then quality matching returns the most relevant ones that meet user quality preferences. The personalized source selection procedure allows reducing the number of candidate sources. Thus only the most relevant ones are transmitted to the mediator via the *mediator interface module* to be a part of the rewriting process. The mediator interface is the access point between the mediator and 2P-Med. It has many other roles. Indeed, this module enriches the initial query by adding information extracted from the user profile at the acquisition module level. It captures also the mediator answer and adapts it to user preferences, for example by changing formats, customizing and re-ordering results. Finally this module handles user feedback either explicitly or automatically by analyzing user behavior. User feedback and his interaction history are then transmitted to the acquisition module to extract the update information.

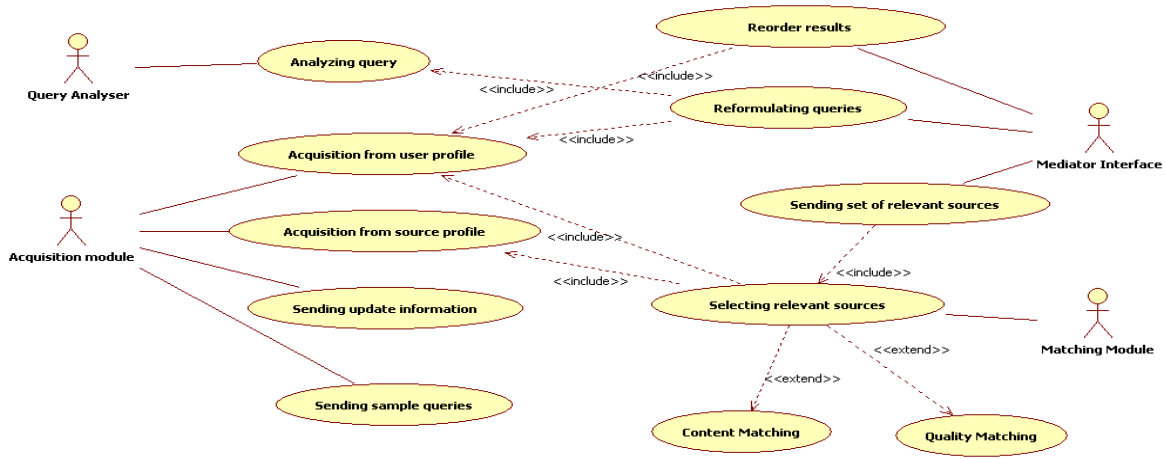

Fig. 6: Use case diagram of the core component

Imane Zaoui et al. / International Journal of Engineering Science and Technology (IJEST)## 4. Case study

We illustrate 2P-Med functionalities with the description of a travel-planning assistant that supports users to plan their trips. Let us consider 7 tourism agencies data sources, tourism office web site, and 2 air company data sources. These heterogeneous and distributed data sources are integrated via a mediator, for example WASSIT [Zellou, (2008)][Wadjinny, (2010)]. User1 is a professor participating to a conference hold in Morocco. So, his objective is *Preparing a holiday in Morocco*, his goals during the interaction session are finding suitable transportation, accommodation, restaurants and other related conferences. He doesn't care about entertainments or monuments.

Traditional mediators give the same response for all users seeking to travel to Morocco if their queries are identical. We use 2P-Med platform to personalize the mediator answers by selecting only the most suitable data sources for each user. We start from describing User1 profile, and then we give an example of two sources profiles. We explain the content matching and the quality matching procedure. The result is a set of relevant data sources. Due to the lack of space, we will not deal with the other personalization techniques like query reformulation, answer customization and user feedback handling.

### 4.1. *User profile*

As we have explained in section 2, user profile is divided into persistent profile and session profile. Each part contains several dimensions. The multidimensional model allows instantiating the profile according to the application needs. In our example, only user goals and user quality preferences are necessary to personalize the mediator answers. We suppose that users set their quality preferences priorities based on the following scale: {0.4: mandatory, 0.3: desirable, 0.2: not desirable, 0.1: indifferent}. Suppose also that User 1 requires the following quality criteria: i) reputation measured using the *Global_Reputation_Score* and completeness measured by *Completeness_Score.* He is indifferent among the other quality factors. For more details about the quality model we propose, refer to [Wadjinny *et al*, (2011)]. User1 quality preferences are $Q_1$:*Global_Reputation_Score>3,* this criterion is mandatory and $Q_3$:*Completeness_Score>30%,* this criterion is desirable. In the following, we present User1 profile. For the indifferent criterion $Q_2$ and $Q_4$, the user quality preference is not specified (NS).

Table 1: User1 session profile (goals and quality preferences)

| Dimension | User1 |
|---|---|
| Goals | ($W_1$:transportation, 0.9; $W_2$:accommodation, 0.6; $W_3$:restaurants, $W_4$:0.7; conference, 0.8) |
| Quality preferences | ($Q_1$, 0.4, >3; $Q_2$, 0.1, NS ; $Q_3$, 0.3, >30%; $Q_4$, 0.1, NS) |

### 4.2. *Sources profiles*

We describe the sources in this case study through the source profile presented in section 2. Remind that the source profile contains four dimensions, which are source identity, source content, source quality characteristics and source ontology. In the tourism domain, many ontologies have been proposed like Quall-Me ontology [Ou *et al*, (2008)], OWL ontology for E-tourism [Cardoso, (2006)], Hi-touch ontology [Mondeca, (2004)] and so on. Our sources could use any of them. In the Table 2 bellow we present the sources profiles of the tourism office portal and a source of a tourism agency. Note that values are illustrative.

Table 2: Source1 and source2 profiles

| | Source 1 | Source 2 |
|---|---|---|
| **Identity** | Id=1, Name , TOURISM Portal<br>URL:www.tourisme.gov.ma<br>Owner : Moroccan Tourism Ministry<br>Type: documents | Id=2, Name: Tourism Agency<br>URL:http://TAgency.com<br>Owner: BestTrip Agency<br>Type: documents/videos |
| **Content** | (Holidays , 0,7; Restaurants , 0.8;<br>Transport , 0.5; Monuments , 0.4, Tourists guides, 0.3) | (Cities , 0.9; Monuments , O.8;<br>Transport , 0.6; Entertainments , 0.4) |
| **Quality** | Reputation= 5<br>Freshness= 1 year<br>Completeness= 70%<br>Time of response= 1s | Reputation= 2<br>Freshness= 5 years<br>Completeness= 20%<br>Time of response= 3s |

### 4.3. *Sources profiles*

Using source content dimension, we select the relevant sources according to User1 goals. As we mentioned before, content matching is based on calculating a similarity score between the content vector and the goals vector. Nevertheless, the vectors contains different key words, we need first to homogenize them. To overcome the problem of vector homogeneity, we propose to take into account the common terms with their relative weights and add absent concepts in each source content vector with a weight of zero. Then we calculate the similarity score as the average of cosine angle, Jaccard and Dice coefficients (Eq. (5)). The threshold defined by User1 is 50%. Results are presented in Table 3.

Table 3: Content matching procedure

|  | $W_1$ | $W_2$ | $W_3$ | $W_4$ | Sim | R |
|---|---|---|---|---|---|---|
| **User1** | 0.9 | 0.8 | 0.7 | 0.6 | | |
| $S_1$ | 0.5 | 0 | 0.8 | 0 | 0.596 | 4 |
| $S_2$ | 0.6 | 0 | 0 | 0 | 0.433 | 6 |
| $S_3$ | 0.2 | 0.6 | 0.4 | 0 | 0.650 | 2 |
| $S_4$ | 0 | 0.7 | 0 | 0.1 | 0.430 | 7 |
| $S_5$ | 0 | 0.3 | 0 | 0 | 0.276 | 8 |
| $S_6$ | 0.8 | 0 | 0.2 | 0.1 | 0.593 | 5 |
| $S_7$ | 0 | 0 | 0 | 0.6 | 0.270 | 9 |
| $S_8$ | 0.2 | 0.4 | 0.3 | 0.1 | 0.643 | 3 |
| $S_9$ | 0.7 | 0.6 | 0.5 | 0.3 | 0.933 | 1 |
| $S_{10}$ | 0 | 0 | 0 | 0 | 0 | 10 |

Table 3 gives the ordered set of selected sources having a score more than 50%, which is {$S_9$, $S_3$, $S_8$, $S_1$, $S_6$}. These selected sources have different quality parameters summarized in Table 4.

Table 4: Quality characteristics of sources

|  | $Q_1$ (score) | $Q_2$ (years) | $Q_3$ (%) | $Q_4$ (s) |
|---|---|---|---|---|
| $S_1$ | 5 | 20 | 50 | 1 |
| $S_3$ | 5 | 30 | 80 | 1 |
| $S_6$ | 3 | 2 | 60 | 0.5 |
| $S_8$ | 4 | 5 | 10 | 2 |
| $S_9$ | 1 | 10 | 20 | 1 |

We apply our algorithm based on calculating a SAW score to select only sources respecting User1 quality preferences. The remaining sources with their SAW scores are given in Table 5.

Table 5: Quality matching procedure

|  | $Q_1$ (score) | $Q_2$ (years) | $Q_3$ (%) | $Q_4$ (s) | Source Score |
|---|---|---|---|---|---|
| $S_1$ | 1 | 0.642 | 0 | 1 | 0.5642 |
| $S_3$ | 1 | 1 | 1 | 1 | 0.9 |
| $S_6$ | 0 | 0 | 0.333 | 0 | 0.0999 |

Sources scores give the following ranking: $S_3$ is more appreciated than $S_1$ and finally $S_6$. As shown in this example, our source selection and ranking algorithm returns to User1 a set of relevant sources that satisfies his profile.

### 5. Conclusion

In this paper, we outline the architecture of our personalization platform for mediation systems (2P-Med) and describe its functionalities, in particular, user modeling, source modeling and source selection. 2P-Med has many advantages. First, the platform is independent from application domains. It could be used for traveling domain as demonstrated in the case study, but also in digital libraries, e-commerce, e-government…etc. Second, 2P-Med is extensible. Indeed, it is easy to add new functionalities and new features as needed. This is due to the modularity of the platform conception and the multidimensionality of the profiles. Third, 2P-Med allows personalizing the mediator responses following a personalization process. Last but not least, among a considerable number of candidate sources, 2P-Med reduces the set of integrated sources via the personalized source selection procedure. This has a great impact on user satisfaction and mediator performances as shown in [Wadjinny, (2010)]. Our platform could be easily plugged in any mediation system. In future works, we plan to use 2P-Med to personalize web services discovering and composition. We will also perform test beds using WASSIT [Zellou, (2008)], the mediator framework developed by our laboratory.